\documentclass{epl}
\input epsf.sty
\def\qn{x}
\def\phin{q}
\def\lfrac#1#2{#1/#2}
\def\db{\,\, {\bar{} \!\!d}\!\,\hspace{0,5pt}}
\def\comment#1{}
\def\lfrac#1#2{#1/#2}

\title{Integrals over Products of Distributions and
  Coordinate Independence of Zero-Temperature Path Integrals}
\shorttitle{Integrals over Products of Distributions\dots
}
\author{H.~Kleinert\thanks{E-mail: kleinert@physik.fu-berlin.de}\inst{} and
     A.~Chervyakov\thanks{E-mail: chervyak@physik.fu-berlin.de.
On leave from LCTA, JINR, Dubna, Russia.
                   }\inst{}}
                     \institute{ Freie Universit\"at Berlin\\
          Institut f\"ur Theoretische Physik\\
          Arnimallee 14, D-14195 Berlin}
\pacs{3.65.-w}{Quantum mechanics}
\pacs{2.90.+p}{Other topics in mathematical methods in physics}
\pacs{2.30.Qy}{Integral transforms and operational calculus}

\begin{document}
\maketitle
\begin{abstract}
In perturbative calculations
of quantum-statistical zero-temperature
 path integrals in
curvilinear coordinates
one encounters Feynman
diagrams involving
multiple temporal integrals over products of distributions,
which are
mathematically undefined.
In addition, there are
terms proportional to powers of Dirac $ \delta $-functions at the origin
coming from the measure of path integration.
We give simple rules
for integrating products of
distributions in such a way that
the results ensure
coordinate independence of
the path integrals.
The rules are derived
by using equations of motion and
partial integration, while
keeping track of certain minimal features
originating in the
unique definition of all singular integrals
in $1 - \epsilon$ dimensions.
Our rules
yield the same results as
the much more cumbersome calculations
in $1- \epsilon$
dimensions where the limit $ \epsilon \rightarrow 0$
is taken at the end.
They also agree with the rules found in an
independent treatment on a finite time interval.
\end{abstract}
\section{Introduction}
While quantum mechanical path integrals in curvilinear coordinates
have  been defined  uniquely
and independently of the choice of coordinates
within the time-sliced formalism
\cite{1}, a perturbative definition
on a continuous time axis
poses severe problems
which have been solved only recently
\cite{2,3}.
To exhibit the origin of the difficulties, consider the
associated partition function
calculated for periodic paths on the imaginary-time axis $\tau $:
\begin{equation}
Z = \int  {\cal D} \phin(\tau ) \, \sqrt{g(q)} \,
e^{-{\cal A} [\phin ]},
\label{@1}\end{equation}
where ${\cal A} [\phin ]$ is the euclidean action
with the general form
\begin{equation}
{\cal A} [\phin ]=\int_0^ \beta  d\tau \left[ \frac{1}{2}\,g_{\mu \nu }(q(\tau ))
\dot q^\mu(\tau )
\dot q^\nu(\tau )+V(q(\tau ))\right].
\label{@}\end{equation}
The dots denote $\tau $-derivatives, $g_{\mu \nu }(q)$ is
a metric, and $g=\det g$
its determinant.
The path integral is
 defined perturbatively as follows:
The metric
$g_{\mu \nu }(q)$
and the potential $V(q)$
 are expanded
around some point $q_0^\mu$
in powers of $ \delta q^\mu\equiv q
^\mu-q_0^\mu$.
After this, the action ${\cal A} [\phin ]$
is separated into a free  part
$
{\cal A}_0[q_0; \delta \phin ]\equiv\int_0^ \beta d\tau \left[
\frac{1}{2}g_{\mu \nu }(q_0)
\partial _t  \delta q^\mu\partial _t \delta  q^\nu+\frac{1}{2} \omega ^2 \delta q^\mu \delta q^ \nu\right]
$,
and an interacting part
${\cal A}_{\rm int}[q_0; \delta \phin ]\equiv
{\cal A}[ \phin ]-
{\cal A}_0[q_0; \delta \phin ]$.

A first problem is encountered
in the measure
of functional integration
in
(\ref{@1}).  Taking $ \sqrt{g(q)}$ into the exponent and expanding
in powers of $ \delta q$, we define
an effective action $ {\cal A}_{ \sqrt{g} }=-\frac{1}{2} \delta (0)
\int_0^ \beta d\tau \log [ g(q_0+ \delta q)/g(q_0)]$
which contains the infinite quantity $ \delta (0)$, the
$ \delta $-function at the origin.
It is a formal representation of
the inverse infinitesimal lattice spacing on the time axis, and is
equal to the linearly divergent
momentum
integral
$\int d p/(2\pi)$.

The second problem arises in the
expansion of $Z$ in powers of the interaction.
Performing all Wick contractions,
$Z$ is expressed as
a sum of loop diagrams.
There are interaction terms involving $ \delta \dot q^2  \delta q^n$
which lead to Feynman integrals
over products of distributions.
The diagrams contain
three types of lines representing
the correlation functions
\begin{eqnarray}
\Delta (\tau -\tau ')&\equiv& \langle  \delta \phin (\tau ) \delta \phin (\tau ')\rangle=
\hspace{0mm}\raisebox{-1mm}{\mbox{\input 1.tps }} ,~\label{p1}\\
\partial _\tau \Delta(\tau -\tau ')&\equiv &\langle \delta  \dot \phin (\tau ) \delta \phin (\tau ')\rangle
=\hspace{0mm}\raisebox{-1mm}{\mbox{\input 3o.tps }} ,~\label{p2}\\
 \partial_ \tau \partial_{\tau '}\Delta (\tau-\tau ')&\equiv&
\langle  \delta \dot \phin (\tau )  \delta \dot\phin (\tau ')\rangle
=\hspace{0mm}\raisebox{-1mm}{\mbox{\input 2.tps }}.~\label{p3}
\label{@}\end{eqnarray}
The right-hand sides define the line symbols
to be used in Feynman diagrams to follow below.

Explictly,
the first correlation function reads
\begin{equation}
 \Delta (\tau ,\tau ')=\frac{1}{2\omega}e^{-\omega|\tau -\tau '|}.
\label{p4}\end{equation}
The second correlation function
has a discontinuity
\begin{equation}
\partial _\tau \Delta(\tau ,\tau ') =
     - \frac{1}{2} \epsilon (\tau - \tau ') e^{-\omega|\tau -\tau '|} ,
\label{p5}\end{equation}
where
\begin{equation}
\epsilon (\tau - \tau ')\equiv-1+ 2\int_{-\infty}^\tau  d\tau''  \delta (\tau'' -\tau ')
\label{p6}\end{equation}
is a distribution
which vanishes at the origin and is equal to $\pm1$ for
positive and negative arguments, respectively.
The third correlation function  contains a
$ \delta $-function:
\begin{equation}
 \partial_ \tau \partial_{\tau '}\Delta (\tau, \tau ') =
  \delta(\tau -\tau ') - \frac{\omega}{2}e^{-\omega|\tau -\tau '|} ,
\label{p7}\end{equation}
Mathematically, the temporal
integrals  over products of such distributions
are undefined \cite{qft}.
In this paper we specify
 these integrals
by imposing the natural requirement
of coordinate independence of
the path integral (\ref{@1}).
By the perturbative calculation
up to three loops we show
that this requirement alone can
not fix uniquely values of all ambiguous
integrals over products of distributions.
However we define a simple consistent
procedure for calculating singular
Feynman integrals. All results obtained
in this way ensure coordinate independence.
They agree with what we have obtained
in our previous work.
In Ref.~\cite{2},
we have shown that Feynman integrals in {\em momentum
space\/} can be uniquely defined
as $ \epsilon \rightarrow 0$ -limits of
$1- \epsilon $ -dimensional
integrals via an  analytic  continuation
\`a la
't~Hooft and Veltman \cite{4}.
This definition makes
path integrals coordinate-independent.
In Ref.~\cite{3} we have given rules
for calculating the same results directly
from Feynman integrals in a $1- \epsilon $ -dimensional
space.

The  calculation procedure developed
in this paper
avoids the cumbersome evaluation of Feynman integrals
in $1- \epsilon $ dimensions. In fact,
it does not require specifying any regularization scheme.
As a fundamental byproduct,
it lays the foundation
for a new
extension of the theory of distributions,
in which also integrals over products
are defined, not only linear combinations.

\section{Perturbation Expansion}
The relations between singular
Feynman integrals will be derived
form the requirement of
coordinate independence
of the
exactly solvable
path integral
of a point particle of unit mass in a harmonic
potential $\omega^2 \qn ^2/2$,
whose action is
\begin{equation}
{\cal A}_ \omega =
\frac{1}{2}\,\int\,d \tau \left[\dot \qn ^2(\tau ) +
\omega^2 \qn ^2(\tau )\right].
\label{m1}\end{equation}
For a large imaginary-time interval $ \beta $,
the partition function is given by the path integral
\begin{equation}
  Z_ \omega  = \int  {\cal D} \qn  (\tau)\,
e^{-{\cal A}_{ \omega } [\qn ]}
= e^{-(1/2) \rm Tr \log (-\partial^2 + \omega^2)} =
  e^{-\beta\lfrac{\omega}{2}}.
\label{mq2}\end{equation}
For simplicity here and in the following
the target space
is assumed to be one-dimensional.
In a higher dimensional euclidean space the problem
of coordinate independence of the path integral (\ref{mq2})
was first examined in the papers \cite{5} and \cite{6}
where different noncovariant quantum corrections
to the classical action (\ref{m1}) were found.
When this space being transformed
holonomically to the curvilinear coordinates
all covariant structures, such as the scalar
curvature, remain, however, zero. Therefore the same
problem can be reexamined directly in one
dimension without loss of
generality.

 A coordinate transformation
turns (\ref{mq2})
into a path integral of the type (\ref{@1})
with a
singular perturbation expansion.
From our work in Refs.~ \cite{2,3}
we know that all
terms in this expansion vanish in dimensional regularization,
thus ensuring the coordinate independence
of the perturbatively defined
path integral. In this paper, we proceed in the opposite direction:
we {\em require\/} the vanishing
of all expansion terms
to find the relations
for integrals over products of distributions.

For simplicity we assume the coordinate transformation
to preserve the symmetry $\qn \leftrightarrow -\qn $ of the initial oscillator,
such that its power series expansion starts out like
$\qn  (\tau)=f(\phin (\tau)) = \phin  - {g}\phin ^3/3 + {g^2}a\phin ^5/5 - \cdots~$,
where $g$ is a smallness parameter, and $a$ some extra parameter.
We shall see that the identities are independent of $a$, such that $a$
will merely
serve
to check the calculations.
The transformation changes the partition function
(\ref{mq2})
into
\begin{equation}
  Z = \int  {\cal D} \phin  (\tau)\,
e^{-{\cal A}_{J} [\phin ]}
e^{-{\cal A} [\phin ]},
\label{m2}\end{equation}
where is $ {\cal A}_{} [\phin ]$ is the transformed action,
whereas
 $ {\cal A}_J[\phin ]$
an effective action coming from the
Jacobian of the coordinate transformation:
\begin{equation}
 {\cal A}_J[\phin ]=
-\delta (0)\int\, d \tau\,\log \,\frac{\delta f(\phin (\tau))}{\delta \phin (\tau)}.
\label{m4}\end{equation}
The transformed action is decomposed into a free
part
\begin{equation}
    {\cal A}_{ \omega } [\phin ]= \frac{1}{2}\,\int\, d \tau [\dot \phin ^2(\tau )
+\omega^2 \phin ^2 (\tau)] ,
\label{m6}\end{equation}
and an interacting part, which reads to
second order in  $g$:
\begin{eqnarray}
\!\!~
{\cal A}_{\rm int} [\phin ] &=& \frac{1}{2} \int\, d \tau
\Bigg\{-g\left[ 2\dot \phin ^2  (\tau) \phin ^2 (\tau)
+ \frac{2\omega^2}{3}\phin ^4 (\tau) \right]
\nonumber\\
& +& g^2
\left[\left(1 + 2a\right) \dot \phin ^2 (\tau)  \phin ^4 (\tau)
+ \omega^2 \left(\frac{1}{9} + \frac{2a}{5}\right)
\phin ^6  (\tau)\right]\Bigg\} . \label{m7}
\end{eqnarray}
To the same order in $g$,
the Jacobian action (\ref{m4}) is
\begin{eqnarray}
 &&\!\! {\cal A}_J[\phin ]=
-\delta (0)\int\, d \tau\left[-g \phin ^2(\tau) +
g^2 \left(a - \frac{1}{2}\right) \phin ^4 (\tau)
\right] . \label{m8}
\end{eqnarray}
For $g=0$,
the transformed partition function
(\ref{m2}) coincides, of course, with (\ref{mq2}).
When expanding   $Z$ of Eq.~(\ref{m2})
in powers of $g$,
we obtain
Feynman integrals
to each order in $g$, whose sum
must vanish
to ensure coordinate independence.
By considering only
connected Feynman diagrams,
we study directly the ground state energy.

\section{Ground State Energy}
Here the coordinate independence
will be tested perturbatively
up to three loops.
The graphical expansion for
the ground state energy
has the following structure:
To given
order $g^n$, there exist
Feynman diagrams with $L=n+1, n,$ and $n-1$
number of loops
coming from the interaction
terms (\ref{m7}) and (\ref{m8}), respectively.
The diagrams are composed of the
three  line
types  in (\ref{p1})--(\ref{p3}), and
new interaction vertices
arising for each
power of $g$. The diagrams coming from the Jacobian action
 (\ref{m8}) are easily recognized by  accompanying factors $ \delta^n (0)$.

To first order in $g$, there
exists only three diagrams, two
originated from the
interaction (\ref{m7}), and one
from the Jacobian action (\ref{m8}):
\begin{equation}
-\,g\hspace{0mm}\raisebox{-1mm}{\mbox{\input 6.tps }} - g\,\omega^{2}\hspace{-27mm}\raisebox{-11.57mm}{\mbox{\input inf.tps }} ~~~~~~~~~~~~~~
~~~
~~~
~~~
 + g\,\delta (0) \hspace{0mm}\raisebox{-1mm}{\mbox{\input 0dot.tps }} .
\label{f1}\end{equation}
~\\[-1.2cm]
To order $g^2$,
we distinguish several contributions.
First there are two three-loop local diagrams
coming
from the interaction (\ref{m7}),
and one two-loop local diagram
from the Jacobian action (\ref{m8}):
\begin{eqnarray}
 &&  g^2\,\Bigg[\,\,\,
 3 \left(\frac{1}{2} + a\right)\hspace{0mm}\raisebox{-3.2mm}{\mbox{\input 7.tps }} +
\,15 \omega^{2} \left(\frac{1}{18} + \frac{a}{5}\right)
 \hspace{-27mm}\raisebox{-13.7mm}{\mbox{\input clover.tps }}
~~~~~~~~~~~~~~~~~~~
~~~
- 3\left(a - \frac{1}{2}\right)\, \delta (0)\,\,
\hspace{-27mm}\raisebox{-11.57mm}{\mbox{\input inf.tps }} ~~~~~~~~~~~~~~
~~~~~~~~~
\,\,\, \Bigg]\,.
\label{f2}\end{eqnarray}
~\\[-1.2cm]
We call a diagram {\em local\/} if it involves
no temporal integral.
The Jacobian
action (\ref{m8}) contributes further
the nonlocal diagrams:
\begin{eqnarray}
 &&\!\!
-\frac{g^2}{2!}\bigg\{
 2\delta^2 (0) \!\!\hspace{0mm}\raisebox{-1mm}{\mbox{\input 0dotdot.tps }}
 \!\!- 4\delta (0) \big[\!\!
\hspace{0mm}\raisebox{-1mm}{\mbox{\input 6dot.tps }}\!+\!\!\!\!\!
\hspace{0mm}\raisebox{-1mm}{\mbox{\input 6pdot.tps }}
\!\!+ 2\,\omega^{4}\!\!\hspace{0mm}\raisebox{-1mm}{\mbox{\input infdot.tps }}\big]\bigg\}.
\label{f3}\end{eqnarray}
The remaining diagrams
come from the interaction  (\ref{m7}) only.
They  are either of the
three-bubble type, or of the watermelon
type, each with all possible combinations
of the three line types (\ref{p1})--(\ref{p3}):
The sum of all three-bubbles diagrams
is
\begin{eqnarray}
&& -\frac{g^2}{2!}\big[~~ 4\hspace{0mm}\raisebox{-1.2mm}{\mbox{\input 8.tps }}
\!+\,\,2~\hspace{0mm}\raisebox{-1.2mm}{\mbox{\input 9.tps }}
~+\,\,\,2\,\,\,\hspace{0mm}\raisebox{-1.2mm}{\mbox{\input 10.tps }}
\nonumber \\
&&~~~+
 ~8\,\omega^2\!
 \hspace{-27.0mm}\raisebox{-11.5mm}{\mbox{\input threeb1.tps }}~~~
~~~~~~~~~~~~~~~~~~~
+ 8\omega^2 \!\!\hspace{-27.0mm}\raisebox{-11.5mm}{\mbox{\input threeb2.tps }}~~~~~~~~~~~~~~~~~~~~~~~
 + 8 \omega^4 \hspace{-27.0mm}\raisebox{-11.5mm}{\mbox{\input threeb.tps }}~~~~~
\quad\quad\quad\quad\quad \quad \big] ~.
\label{f4}\end{eqnarray}
~\\[-1.2cm]
The watermelon-like diagrams
contribute
\begin{eqnarray}
& &
{\!\!-\frac{g^2}{2!}\, 4 \,\bigg[\!\!
\!\!\!\hspace{0mm}\raisebox{-2mm}{\mbox{\input 11.tps }}
\!\!+ 4\!\!\!\hspace{0mm}\raisebox{-1.95mm}{\mbox{\input 12.tps }}
\!\!+\!\!\!\!\! \hspace{0mm}\raisebox{-1.9mm}{\mbox{\input 13.tps }}
\!\!+  4\omega^2\! \hspace{-27.0mm}\raisebox{-12.3mm}{\mbox{\input waterm2.tps }}
~~~~~~~~~~~
~~~~~~~~~~~
 \!+\! \!\frac{2}{3}\omega^4 \!
\hspace{-27.0mm}\raisebox{-12.3mm}{\mbox{\input waterm.tps }}
~~~~~~~~~~~
~~~~~~~~~~~
\bigg].   \!\!\!\!\!        }
\label{f5}
\end{eqnarray}
~\\[-01.1cm] Since the equal-time
expectation value $\langle\dot\phin (\tau)\,\phin (\tau)\rangle$
vanishes according to Eq.~(\ref{p5})
there are, in addition, a number of trivially vanishing diagrams,
 which have been omitted.

In our
previous work \cite{2,3},
all
integrals were calculated
in
$D=1-\varepsilon$ dimensions,
taking the limit
$\varepsilon\rightarrow 0$ at the end.
In this way we confirmed
that the sums of all
Feynman diagrams
contributing to each order in $g$ vanish.
Here we proceed in the opposite direction and
derive the rules for integrating products of distributions
from the vanishing of the sums.

\section{Imposing Coordinate Independence}
In a first step
we simplify the above perturbation
expansion of the ground state
energy using
the inhomogeneous field
equation satisfied by the correlation  function
(\ref{p4}):
\begin{equation}
\ddot \Delta  (\tau) = - \int \db k\, \frac{k^2}{k^2+ \omega^2}
 e^{ik\tau} = - \, \delta (\tau) + \omega^2  \Delta (\tau)\,,
\label{i1}\end{equation}
and the following
well-defined integrals
for two correlation  functions:
\begin{equation}
\!\! \int d \tau\,\left[\dot\Delta^2 (\tau)
+ \omega^2  \Delta ^2 (\tau)\right] =
   \Delta (0),
\label{i2}\end{equation}
as well as the
integrals containing
four correlation  functions:
\begin{eqnarray}
\int d \tau \,\Delta ^4 (\tau) &=& \frac{1}{4\omega^2}\,\Delta ^3 (0) ,
\label{i5} \\
\int d \tau\, \dot\Delta ^2 (\tau) \Delta ^2 (\tau) &=&
\frac{1}{4}\,\Delta ^3 (0) ,\label{ibis}\\
 \int d \tau\,\dot\Delta^4 (\tau) &=&
 \frac{1}{4}\,\omega^2 \,\Delta ^3 (0) .
\label{i10}\end{eqnarray}
These integrals are
obtained directly by substituting
the explicit representations
(\ref{p4}) and (\ref{p5}) into (\ref{i2}),
(\ref{i5}), (\ref{ibis}) and (\ref{i10}),
respectively.
Also we shall use
the singular integral
obtained by the same
direct substitution
of two correlation functions:
\begin{eqnarray}
&&
\int\! d \tau \!\left[\ddot\Delta^2 (\tau)
+ 2\omega^2 \dot \Delta ^2 (\tau)
+\omega^4  \Delta ^2 (\tau)
\right]
 \!=\! \int d \tau \,
\delta^2 (\tau)\,,
\label{i3}\end{eqnarray}
where the last integral containing
the square of the $\delta$-function
will be specified later by the requirement
of coordinate independence.

Consider now the perturbation
expansion of the ground state
energy.
To first order in $g$,
the sum of
Feynman diagrams (\ref{f1})
must vanish:
\begin{equation}
\hspace{0mm}\raisebox{-1mm}{\mbox{\input 6.tps }} + \,\omega^{2}\hspace{-27mm}\raisebox{-11.57mm}{\mbox{\input inf.tps }} ~~~~~~~~~~~~~~
~~~
~~~
~~~
 -\,\delta (0) \hspace{0mm}\raisebox{-1mm}{\mbox{\input 0dot.tps }} = 0 .
\label{di1}\end{equation}
~\\[-1.2cm]
The analytic form of this relation is
\begin{equation}
\left[
- \ddot \Delta  (0) + \omega^2  \Delta (0)
  - \, \delta (0)\right]\,\Delta (0) = 0 ,
\label{ai1}\end{equation}
the zero on the right-hand side being
a direct consequence
of the equation  of motion (\ref{i1})
for the correlation  function
at origin.

To order $g^2$,
the same equation
reduces the sum of all local diagrams in
(\ref{f2}) to a finite result plus a
term proportional to $ \delta (0)$:
\begin{eqnarray}
&& \!\!\!\!\!\!\!\left[\!- 3 \left(\frac{1}{2} \!+\! a \right)
 \ddot\Delta (0) \!+\!
15 \left(\frac{1}{18} + \frac{a}{5}\right)
\omega^2 \Delta (0)
- 3 \left(a \!-\! \frac{1}{2}\right)\delta (0)
\right]\!\!\Delta^2 (0)
 \!=\! \left[3 \delta (0)
 \!-\! \frac{2}{3} \,\omega^2 \Delta (0)\right]\!\!\Delta^2 (0).
\nonumber
\label{ai4}\end{eqnarray}
Representing  right-hand side diagrammatically, we obtain
the identity
\begin{equation}
\!\!\!\!\mathop{\Sigma }{(\ref{f2})} =
  3 \delta (0)\hspace{-27mm}\raisebox{-11.57mm}{\mbox{\input inf.tps }} ~~~~~~~~~~~~~~
~~~~~~~~~
- \,\frac{2}{3}\, \omega^{2}
 \hspace{-27mm}\raisebox{-13.7mm}{\mbox{\input clover.tps }}~~~~~~~~~~~~~~~~~~~~~~ ,
\label{di4}\end{equation}
~\\[-1.4cm]
where
$\mathop{\Sigma }{(\ref{f2})}$ denotes the sum of all diagrams in Eq.~(\ref{f2}).
 Using (\ref{i2})
together with the field equation (\ref{i1}),
we reduce the sum (\ref{f3})
of all one and two-loop bubbles
diagrams to terms involving $ \delta (0)$ and $ \delta ^2(0)$:
\begin{eqnarray}
&& \!\!\!\!\!\!\!\!\!\!\!\!\!-  \frac{2}{2!}
\left\{\delta^2 (0) \int \!\!d \tau\,\Delta^2 (\tau)
 - 2\delta (0)\!\! \int \!\!d \tau\,\left[
\Delta (0)\dot\Delta^2 (\tau)
\!-\! \ddot\Delta (0) \Delta^2 (\tau)
 \!+\! 2\omega^2 \Delta (0)\Delta^2 (\tau)\right]\right\}
 \nonumber\\
& &~~~~~~~~~~~~~~~~~~=  2\delta (0)\,\Delta^2 (0)
 +  \delta^2 (0)\int d \tau \Delta^2 (\tau) .
\label{ai2}\end{eqnarray}
Hence we find  the diagrammatic identity
\begin{equation}
\mathop{\Sigma }{(\ref{f3})} =
 2\delta (0)\hspace{-27mm}\raisebox{-11.57mm}{\mbox{\input inf.tps }} ~~~~~~~~~~~~~~
~~~~~~~~~
+\,\, \delta^2 (0) \!\!\hspace{0mm}\raisebox{-1mm}{\mbox{\input 0dotdot.tps }} .
\label{di2}\end{equation}
~\\[-1.2cm]
Now, the terms accompanying $\delta^2 (0)$
turn out
to  cancel
similar terms coming from the sum
of all three-loop bubbles
diagrams in (\ref{f4}).
In fact, the
relations (\ref{i2}) and (\ref{i3})
lead to
\begin{eqnarray}
&&\!\!\!\!\!\!\!\!
- \frac{2}{2!}\int d \tau\,\left[
- 2\Delta (0)\ddot\Delta (0)\dot\Delta^2 (\tau)
+ \Delta^2 (0)\ddot\Delta^2 (\tau)
+ \ddot\Delta^2 (0)\Delta^2 (\tau)
+ 4\omega^2 \Delta^2 (0)\dot\Delta ^2 (\tau)
\right.\nonumber\\&  &\!\!\!\!\!\!\!\! \left.
\!- 4\omega^2 \Delta (0)\ddot\Delta (0)\Delta^2 (\tau)
 \!+\!4\omega^4 \Delta^2 (0) \Delta^2 (\tau)
\right]
\!\!= -\left[\int d \tau\,\delta^2 (\tau) +  2 \delta (0)\right]\! \Delta^2 (0)
 \!- \delta^2 (0)\!\int d \tau \, \Delta^2 (\tau) . \nonumber
\label{ai3}\end{eqnarray}
Thus we find the diagrammatic
identity for all bubbles diagrams
\begin{equation}
\mathop{\Sigma }{(\ref{f3})}
 + \mathop{\Sigma}{(\ref{f4})} =
 - \int d \tau\,\delta^2 (\tau)\hspace{-27mm}\raisebox{-11.57mm}{\mbox{\input inf.tps }} ~~~~~~~~~~~~~~
~~~~~~~~~\,.
\label{di3}\end{equation}
~\\[-1.2cm]

The first two
watermelon
diagrams in Eq.~(\ref{f5})
correspond to the
integrals
\begin{eqnarray}
I_1 &=& \int d \tau\,\ddot\Delta ^2 (\tau) \Delta ^2 (\tau)\,,
\label{i66}\\
I_2 &=&\int d \tau \,\ddot\Delta (\tau) \dot\Delta ^2 (\tau) \Delta (\tau),
\label{i7}
\end{eqnarray}
whose evaluation is subtle. Consider first the integral
(\ref{i66}) which
contains a
square of a $\delta$-function. We separate this out
by writing
\begin{eqnarray}
&& I_1 = \int d \tau\,\ddot\Delta ^2 (\tau) \Delta ^2 (\tau)
 =  I_{1}^{\rm{div}}
+ I_{1}^{R} ,
\label{i666}\end{eqnarray}
with a divergent and a regular part
\begin{eqnarray}
I_{1}^{\rm{div}}= \Delta^2 (0)\,\int d \tau\,\delta ^2 (\tau)\,,
 ~~~~~I_{1}^{R} =  \int d \tau\,
\Delta^2 (\tau)\left[\ddot\Delta ^2 (\tau) - \delta^2 (\tau)\right].
\label{i666b}\end{eqnarray}
All other watermelon diagrams (\ref{f5})
lead to the well-defined
integrals  (\ref{i10}), (\ref{ibis}),
and (\ref{i5}), respectively.
Substituting these and
(\ref{i7}), (\ref{i666})
into (\ref{f5}) yields
the sum of all watermelon
diagrams
\begin{eqnarray}
&  &\!\!\!\!\!\!\!\!\!\!- \frac{4}{2!}\int d \tau\,\left[
\Delta^2 (\tau) \ddot\Delta^2 (\tau)
+4\Delta (\tau) \dot\Delta^2 (\tau) \ddot\Delta (\tau)\phantom{\int}
\!\!\!\!\!\!+\dot\Delta^4 (\tau)
 + 4\omega^2 \Delta^2 (\tau) \dot\Delta ^2 (\tau)
+ \frac{2}{3}\omega^4 \Delta^4 (\tau)
\right] \nonumber \\
&&~~~~= ~- ~2\Delta^2 (0)\,\int d \tau\,\delta^2 (\tau) -
2 \left(I_1^R + 4 I_2  \right)
 - \frac{17}{6}\,\omega^2 \Delta^3 (0)\,.
\label{ai5}\end{eqnarray}
Adding these to
(\ref{di4}), (\ref{di3}),
we obtain the sum of all
second-order connected diagrams
\begin{equation}
\mathop{\Sigma }{(\rm {all})} =
3\left[ \delta (0)
 - \int d \tau\,\delta^2 (\tau)
 \right]\Delta^2 (0)
 - 2\, \left(I_1^R + 4 I_2  \right)
 - \frac{7}{2}\,\omega^2 \Delta^3 (0)\,,
\label{all}\end{equation}
where the integrals $I_1^R$ and $I_2$
are undefined so far.
This sum has to vanish to guarantee
coordinate independence.
We therfore
equate to zero
both the singular and finite
contributions in Eq.~(\ref{all}).
The first
yields the rule
for the
product of two $ \delta $-functions
\begin{equation}
\delta^2 (\tau) = \delta (0)\,\delta(\tau)\,.
\label{i11a}\end{equation}
This equality should of course
be understood
in the distributional sense, it holds after
multiplying it with an arbitrary test function
and integrating over $\tau $.
As it was shown in our previous paper \cite{nbeta},
the rule (\ref{i11a})
guarantees the cancellation
of all short-distance
singularities to any
order of the perturbation theory.
This happens independently
of the value of $ \delta (0)$, making a
regularization superfluous.
There is  perfect cancellation
of all powers of $ \delta (0)$
arising from the expansion  of the Jacobian action,
which
is the fundamental reason why the heuristic
{\em Veltman rule\/}  of setting
$\delta (0)=0$  is applicable everywhere without problems.

The vanishing of the regular parts of (\ref{all})
requires the integrals (\ref{i7})
and (\ref{i666}) to satisfy
\begin{equation}
I_1^R + 4 I_2  = - \frac{7}{4}\,\omega^2 \Delta^3 (0) =
- \frac{7}{32\omega}\,.
\label{i11b}\end{equation}
At this point we run into two difficulties.
First,
this single equation (\ref{i11b})
for the two undefined integrals
$I_1^R$ and $I_2$
is insufficient to
specify both integrals, so that
the requirement of reparametrization
invariance alone is not enough to fix
all ambiguous temporal integrals over
products of distributions.
Second, and more seriously,
Eq.~(\ref{i11b})
leads to conflicts with
standard integration rules based on the use of
partial integration
and equation of motion.
Let us apply these rules to
the calculation of
the integrals $I_1^R$ and $I_2$
in different orders.
Inserting  the equation of motion (\ref{i1})
into the finite part of the integral
(\ref{i666}) and making use of the
regular integral (\ref{i5}),
we find immediately
\begin{eqnarray}
 I_{1}^{R} &=& \int d\tau\,  \Delta^2 (\tau)
  \left[ \ddot\Delta^2 (\tau) -
   \delta^2 (\tau) \right]  \nonumber \\
& =&
- 2\omega^2\,\Delta^3 (0)  + \omega^4\,\int  d\tau \,\Delta ^4  (\tau)
 = - \frac{7}{4}\omega^2 \Delta^3 (0) =   - \frac{7}{32\omega}\,.
\label{1*}\end{eqnarray}
The same substitution
of the equation of motion (\ref{i1}) into the other
ambiguous integral $I_2$ of (\ref{i7})
leads, after performing the
the regular integral (\ref{ibis}),
to
%
\begin{eqnarray}
 I_{2} &=& - \int  d\tau \,
 \dot\Delta^2  (\tau)\, \Delta (\tau)\,\delta (\tau) 
    + \omega^2\, \int  d\tau \,
  \dot\Delta^2 (\tau)\,\Delta^2 (\tau)
\nonumber \\
 & = & - \frac{1}{8\omega}\,\int  d\tau \,
 \epsilon^2 (\tau)\,\delta (\tau)  +
  \frac{1}{4}\omega^2 \Delta^3 (0)
  =  \frac{1}{8\omega} \left(- I + \frac{1}{4}\right)\,,
\label{2*}\end{eqnarray}
where $I$
denotes the
undefined integral
over a product of distributions
\begin{equation}
 I = \int d \tau\,\epsilon^2 (\tau) \delta (\tau)\,.
\label{4*}\end{equation}
This integral can apparently be fixed by
applying partial integration
to the integral (\ref{i7})
which reduces it to the completely
regular form (\ref{i10}):
\begin{eqnarray}
\!\hspace{-1cm} I_{2} & = & \frac{1}{3}
\int d\tau\,\Delta
 (\tau) \frac{d}{d\tau } \left[\dot\Delta^3 (\tau) \right]
=
  - \frac{1}{3} \int d\tau \,\dot\Delta^4 (\tau)
 = - \frac{1}{12}\omega^2 \Delta^3 (0) = - \frac{1}{96\omega}.
\label{3*}\end{eqnarray}
There are no boundary terms due to the
exponential vanishing at infinity
of all functions involved.
From (\ref{2*}) and (\ref{3*})
we conclude that $I = 1/3$.

At this point we observe a conflict.
The results (\ref{3*})
and
(\ref{1*})
do not obey
the equation (\ref{i11b}), such that
they are incompatible
with the necessary coordinate
independence of the path integral.
This was the reason
to add  a noncovariant
quantum correction term
$\Delta V = - g^{2}(q^{2}/6)$ to the
classical action (\ref{m1})
in the previous paper \cite{6}.
For perturbative calculation on a finite-time
interval it was also done in \cite{beta}.

From the perspective
of our previous papers
\cite{2,3} where
all integrals were defined in $d=1- \epsilon $ dimensions
and continued to $ \epsilon \rightarrow 0$ at the end,
the above-observed inconsistency
is obvious:
Arbitrary application
of partial integration
and equation of motion
to one-dimensional
integrals is forbidden
whenever several dots can
correspond  to different
contractions of
partial derivatives $\partial _ \alpha ,\, \partial _  \beta ,
\dots$,
from which they arise in the
limit
$d \rightarrow 1$.
The different contractions may lead
to different integrals.
In the pure one-dimensional
calculation of the integrals
$I_1^R$ and $I_2$ this
ambiguity can be accounted
for by using partial integration and
equation of motion (\ref{i1})
only according to the following
integration rules:

1. We perform a
partial integration
which allows us to
apply subsequently the equation of
motion (\ref{i1}).

2. If  the equation of
motion (\ref{i1}) leads to
integrals of the type (\ref{4*}), they
must be performed using the
naively the Dirac rule for the $\delta$-function
and the property $\epsilon (0) = 0$.

3. The above procedure leaves in general singular integrals,
which must be treated once more with the same rules.

Let us show that
calculating the integrals
$I_1^R$ and $I_2$ with
these
rules is consistent with the coordinate independence
condition (\ref{i11b}).
In the
integral $I_2$ of (\ref{i7})
we first
apply partial integration to find
\begin{eqnarray}
\!\hspace{-1cm} I_{2} & = & \frac{1}{2}
\int d\tau\, \Delta (\tau)\,\dot\Delta
 (\tau) \frac{d}{d\tau } \left[\dot\Delta^2 (\tau) \right] \nonumber \\
\hspace{-1cm}& = &
- \frac{1}{2} \int  d \tau \, \dot\Delta^4 (\tau)
- \frac{1}{2} \int  d \tau \, \Delta (\tau)\,
\dot\Delta^2 (\tau)\, \ddot\Delta (\tau)\,,
\label{5*}\end{eqnarray}
with no contributions from the boundary terms.
Note that the partial integration (\ref{3*})
is forbidden since
it does not allow
for a subsequent application of
the equation of motion (\ref{i1}).
On the right-hand side of (\ref{5*})
it can be applied.
This leads to a combination of two
regular integrals (\ref{ibis})
and (\ref{i10}) and the singular integral $I$, which we evaluate
with the naive Dirac rule to $I=0$, resulting in
\begin{eqnarray}
 I_{2} &=&
- \frac{1}{2} \int  d \tau \, \dot\Delta^4 (\tau)
+ \frac{1}{2} \int  d\tau \,
 \dot\Delta^2  (\tau)\, \Delta (\tau)\,\delta (\tau) 
    -\frac{1}{2}\omega^2\, \int  d\tau \,
  \dot\Delta^2 (\tau)\,\Delta^2 (\tau)
\nonumber \\
 & = & \frac{1}{16\omega}\,I  -
  \frac{1}{4}\omega^2 \Delta^3 (0)
  = - \frac{1}{32\omega} \,.
\label{6*}\end{eqnarray}

If we calculate the finite
part $I_1^R$ of the integral (\ref{i666})
with the new rules we obtain a result different from
(\ref{1*}).
Integrating the first term in
brackets by parts and using
the equation of motion
(\ref{i1}), we obtain
\begin{eqnarray}
 I_{1}^{R} & =& \int  d\tau  \Delta^2 (\tau)
  \left[ \ddot\Delta^2 (\tau) -
   \delta^2 (\tau) \right]  \nonumber \\
&=& \int d\tau \left [- \dot{\phantom{\Delta}}\hspace{-2mm}
\dot{\Delta}\hspace{-2mm}\dot{\phantom{\Delta}}
(\tau)\,\dot\Delta (\tau)\, \Delta^2  (\tau )
 - 2\ddot\Delta(\tau)
\,\dot\Delta^2  (\tau)
\,\Delta (\tau) - \Delta^2 (\tau)\,\delta^2 (\tau) \right ]\nonumber \\
&=& \int d\tau \left [\dot\Delta (\tau)\,
\Delta^2  (\tau)\,\dot\delta (\tau) -
  \Delta^2  (\tau)\,\delta^2 (\tau)\right ]
-
  2I_{2} - \omega^2\,\int d\tau  \dot\Delta^2  (\tau)\,\Delta^2  (\tau)\,.
\label{7*}\end{eqnarray}
The last two terms are
already known, while
the remaining singular integral
in brackets must be
subjected once more to the same treatment.
It is integrated by  parts
so that the equation of motion (\ref{i1})
can be applied to obtain
\begin{eqnarray}
&&\!\!\!\!\!\!\!\!\!\!\!\!\!\!\!\!\!\int d\tau \left [\dot\Delta (\tau)\,
\Delta^2  (\tau)\,\dot\delta (\tau) -
  \Delta^2  (\tau)\,\delta^2 (\tau)\right ]
\nonumber\\
&&
\!\!\!\!\!\!\!\!\!\!\!\!= - \int  d\tau  \left[ \ddot\Delta (\tau) \,
\Delta^2 (\tau)  + 2\dot\Delta^2 (\tau)\,\Delta (\tau) \right] \delta (\tau)
- \int d\tau \Delta^2  (\tau )\,\delta^2 (\tau)
= - \omega^2\,\Delta^3  (0) - \frac{1}{4\omega}\,I \,.
\label{8*}\end{eqnarray}
Inserting this into Eq.~(\ref{7*})
yields
\begin{eqnarray}
 I_{1}^{R} & =& \int  d\tau  \Delta^2 (\tau)
  \left[ \ddot\Delta^2 (\tau) -
   \delta^2 (\tau) \right]
=
 - 2I_{2} - \frac{5}{4}\omega^2\,\Delta^3  (0) - \frac{1}{4\omega}\,I =
 - \frac{3}{32\omega}\,,
\label{9*}\end{eqnarray}
the right-hand side following for $I = 0$.

We see now that the integrals
(\ref{6*}) and (\ref{9*}) calculated with the new rules
obey the equation (\ref{i11b}) which guarantees
coordinate independence
of the path integral.

The applicability of the rules (1-3)
follows immediately
from the previously established
dimensional continuation \cite{2,3}.
It avoids completely
 the cumbersome
calculations in $1-\varepsilon$-dimension
with the subsequent limit $ \varepsilon\rightarrow 0$.
Only some intermediate steps
of the derivation require keeping track
of the
$d$-dimensional origin of the rules.
For this, we continue the imaginary time
coordinate $\tau$ to
a $d$-dimensional spacetime vector
$\tau \rightarrow x^\mu = (\tau , x^1, \dots, x^{d-1})$.
and note that
the equation of motion (\ref{i1}) becomes a scalar
field equation of the Klein-Gordon type
\begin{equation}
\left( - \partial_\alpha ^2 + \omega^2 \right) \Delta (x) = \delta^{(d)} (x)\,.
\label{i1d}\end{equation}
In $d$ dimensions,
the relevant second-order
diagrams are obtained by
decomposing the harmonic expectation value
\begin{equation}
\int d^d x \,<q_{\alpha}^2 (x)\,q^2 (x)\,q_{\beta}^2  (0)\,q^2 (0)>\,
\label{cf}\end{equation}
into a sum of products of four two-point
correlation functions according
to the Wick rule. The fields $q_ \alpha (x)$
are the
$d$-dimensional
extensions  $q_ \alpha (x)\equiv \partial _ \alpha q(x)$
of $\dot q(x)$.
Now the $d$-dimensional integrals,
corresponding to
the integrals (\ref{i66}) and (\ref{i7}),
are defined uniquely
by the contractions
\begin{eqnarray}
I_1^d  & = &
\unitlength1em
\int d^d x \,<
\put(0,1){\line(1,0){8.5}}
\put(0,1){\line(0,-1){0.2}}
\!\!
q_{\alpha} (x)
\phantom{.}\put(0,1.2){\line(1,0){8.4}}
\put(0,1.2){\line(0,-1){0.4}}
\hspace{-.2em}
q_{\alpha} (x)
\phantom{i}\put(0,1.4){\line(1,0){8.2}}
\put(0,1.4){\line(0,-1){0.6}}
\hspace{-.3em}q(x)
\phantom{i}\put(0,1.6){\line(1,0){8.1}}
\put(0,1.6){\line(0,-1){.8}}
\hspace{-.3em}q(x)
\put(.25,1){\line(0,-1){0.2}}
q_{\beta} (0)
\put(0.2,1.2){\line(0,-1){0.4}}
q_{\beta} (0)
\put(0.3,1.4){\line(0,-1){0.6}}
q(0)
\put(0.3,1.6){\line(0,-1){.8}}
q(0)> \nonumber\\
 &=&
\int d^d x \,\Delta^2 (x)\,\Delta_{\alpha \beta}^2 (x)\,,
\label{i66d}\\
I_2^d  &=&
\unitlength1em
\int d^d x \,<
\put(0,1){\line(1,0){8.4}}
\put(0,1){\line(0,-1){0.2}}
\hspace{-.3em}q_{\alpha} (x)
\phantom{.}\put(0,1.2){\line(1,0){10.6}}
\put(0,1.2){\line(0,-1){0.4}}
\hspace{-.3em}q_{\alpha} (x)
\phantom{.}\put(0,1.4){\line(1,0){5.9}}
\put(0,1.4){\line(0,-1){0.6}}
\hspace{-.3em}q(x)
\phantom{i}\put(0,1.6){\line(1,0){8.1}}
\put(0,1.6){\line(0,-1){.8}}
\hspace{-.3em}q(x)
\put(0.25,1){\line(0,-1){0.2}}
q_{\beta} (0)
\put(0.23,1.4){\line(0,-1){0.6}}
q_{\beta} (0)
\put(0.24,1.2){\line(0,-1){0.4}}
q(0)
\put(0.23,1.6){\line(0,-1){.8}}
q(0)>
\nonumber\\
&=&
\int d^d x \,\Delta (x)\,\Delta_{\alpha} (x)\,\Delta_{\beta} (x)\,
\Delta_{\alpha\beta} (x)\,.
\label{i7d}\end{eqnarray}
The different derivatives
$\partial _ \alpha \partial  _\beta $
acting on $\Delta  (x)$
prevent us from applying
the field equation (\ref{i1d}). This
obstacle was hidden in the
one-dimensional formulation.
It can be overcome by a partial integration.
Starting  with $I_2^d$, we obtain
\begin{eqnarray}
I_2^d &=& - \frac{1}{2}
\int d^d x \,\Delta_{\beta}^2 (x)
\left[\Delta_{\alpha}^2  (x) + \Delta (x)\,
\Delta_{\alpha\alpha} (x) \right]\,,
\label{10*}
\end{eqnarray}
Treating $I_1^d$ likewise we find
\begin{eqnarray}
 I_1^d  & = & - 2I_2^d +
 \int d^d x \,\Delta^2 (x)\,\Delta_{\alpha \alpha}^2 (x)
 + 2\int d^d x \,\Delta (x)\,\Delta_{\beta}^2 (x)\,\Delta_{\alpha\alpha} (x)\,.
\label{11*}\end{eqnarray}
In the second equation we have used the fact that
$\partial_{\alpha} \Delta_{\alpha\beta} =
\partial_{\beta} \Delta_{\lambda\lambda}$.
The right-hand sides of
(\ref{10*}) and (\ref{11*})
contain now the contracted derivatives $\partial _ \alpha ^2$
such that we can apply
the field equation (\ref{i1d}).
This mechanism
works to all orders
in the perturbation expansion
which is the reason for the applicability
of the rules (1) and (2)
which led
to the results (\ref{6*}) and (\ref{9*})
ensuring coordinate independence.

The rule (3) is  a consequence
of regularized equation
in $d = 1 - \varepsilon$
dimension
\begin{equation}
\Delta (x)\, \Delta_{\alpha}^2 (x)\,\delta (x) \simeq
 |x|^{2\varepsilon}\,\delta (x) = 0 \,,
\label{12*}\end{equation}
which makes the singular
product of distributions
in Eq.~(\ref{4*}) vanishes
before taking $\varepsilon \rightarrow 0$-limit.
This rule is in agreement with
the earlier result derived for finite time intervals in Ref.~\cite{nbeta}.

Let us illustrate
how
different contractions
of partial derivatives $\partial _ \alpha ,\, \partial _  \beta ,
\dots$, may
lead to different integrals
in the $d \rightarrow 1$-limit.
The simplest example is
the anomalous integral
\begin{equation}
I_{{\rm an}}^d =
\int d^dx \,\Delta^2 (x)\,\Delta_{\alpha\beta }^2 (x) -
\int d^dx \,\Delta^2 (x)\,\Delta_{\alpha\alpha}^2 (x)\,,
\label{13*}\end{equation}
where the integrals on the right-hand side
are indistinguishable in $d=1$ dimension.
However, it follows immediately
from Eqs.~(\ref{10*}) and (\ref{11*})
that
\begin{eqnarray}
I_{{\rm an}}^d
& =& \int d^d x \,\Delta_{\alpha}^2 (x)\,\Delta_{\beta }^2 (x) +
3\int d^d x \,\Delta (x)\,\Delta_{\beta}^2 (x)\,
\Delta_{\alpha\alpha} (x)
\nonumber\\
& =& \int d^d x \,\Delta_{\alpha}^2 (x)\,\Delta_{\beta }^2 (x) +
3\omega^2 \int d^d x \,\Delta^2  (x)\,\Delta_{\beta}^2 (x)
\nonumber\\
& -& 3\int d^d x \,\Delta (x)\,\Delta_{\beta}^2 (x) \delta (x)
\mathop{\longrightarrow }_{d \rightarrow 1}
\frac{1}{8 \omega }.
\label{14*}\end{eqnarray}
The dimensional continuation
of all other second-order diagrams
does not change the results of
one-dimensional calculation.
Only the second diagram in Eq.~(\ref{f4})
requires taking more care.
However, the corresponding
$d$-dimensional integral
has the property
\begin{equation}
 \Delta^2 (0) \int d^dx \,\Delta_{\alpha\beta }^2 (x) =
\Delta^2 (0) \int d^dx \, \Delta_{\alpha\alpha}^2 (x)\,,
\label{15*}\end{equation}
which allows us to apply the equation
of motion (\ref{i1d}) directly
as in one dimension.
Thus, by keeping only track
of a few essential properties
of the theory in $d$ dimensions
we indeed obtain a simple
consistent procedure for calculating singular
Feynman integrals. All
results obtained in this way
ensure coordinate independence.
Our procedure
gives us unique rules telling us where we are allowed
to apply partial integration
and the equation of motion
in one-dimensional expressions.
Ultimately, all integrals
are brought to a regular form, which can be
continued back to one time dimension for a
direct evaluation. This procedure
is obviously  much  simpler than
the previous explicit calculations in $d$-dimension
with the limit $d \rightarrow 1$ taken at the end.

The above rules for integrating products of distributions
are in complete agreement with the rules derived
earlier
in a different way
for path integrals on a finite time interval
where the infrared regulator  $ \omega ^2q^2$ was not needed
in the free part of the action ${\cal A}_0$.

Let us briefly discuss the alternative
possibility of giving up partial integration
completely
in ambiguous integrals containing
$\epsilon$- and
$\delta $-function, or their time
derivatives, which makes
unnecessary to satisfy
Eq.~(\ref{3*}).
This yields a freedom
in the definition of integral
over product of distribution
(\ref{4*}) which can be used
to fix $I=1/4$
from the requirement of
coordinate independence
\cite{7}. Indeed,
this value of $I$ makes the integral (\ref{2*})
equal to $I_2 = 0$
such that (\ref{i11b}) is satisfied and
coordinate independence ensured.
In contrast, giving up
partial integration,
the authors of Refs.~\cite{5,8}
have assumed the vanishing
$ \epsilon^2 (\tau )$ at $\tau =0$
so that the integral $I$
should vanish as well: $I=0$.
Then Eq.~(\ref{2*})
yields $I_2 = 1/32\omega$
which together with
(\ref{1*}) does not
obey the coordinate independence condition (\ref{i11b}), making
yet an another
noncovariant quantum correction
$\Delta V = g^{2}(q^{2}/2)$ necessary in the action,
which we reject since it contradicts Feynman's
original rules of path integration.
We do not consider giving up partial integration
as an attractive option
since it is an important tool
for calculating higher-loop diagrams.

\section{Summary}
We have
 defined
 singular Feynman integrals
in perturbative
calculation of zero-temperature  path integrals
in such a way that
coordinate independence is guaranteed.
To second-order in perturbation theory, one encounters two singular integrals
containing products of distributions
and finds one relation between them
due to coordinate independence.
The evaluation of the individual integrals
 was made unique by analytic regularization
 \cite{2,3}.
In this paper we have shown that there are simple rules
for obtaining the same results
without the analytic continuation
of Feynman integrals to $d$  dimensions.
We have merely
kept track of the different
contractions of the derivatives
and performed partial integrations
until one obtains Laplace operators
which allow us to use
the $d$-dimensional field equation
which removes one singularity.
If repeated
recursively, this procedure
leads to regular integrals.
There is no need to
 specify
a regularization scheme, and
calculations are
 much simpler
than
the previous ones
in $d$-dimension with the limit
$d \rightarrow 1 $ taken at the end.
Our results
are agreement with
the earlier result derived for finite time intervals in Ref.~\cite{nbeta}.

Just as in the time-sliced definition of
path integrals
in curved space in Ref.~\cite{1},
there is absolutely no need for extra compensating potential terms
found necessary in the treatments in Refs.~\cite{5,7,6}.

%
%
%


\end{document}